\documentstyle[aps,preprint,tighten,floats,eqsecnum,epsf,rotate,prl]{revtex}

\newbox\rotbox

\begin{document}

\preprint{\vbox{
\noindent Submitted to {\it Phys.\ Rev.\ D}
\hfill DOE/ER/40427-27-N95\\}}

\title{\huge QCD Equalities\\ for\\ Baryon Current Matrix Elements}

\author{\sc Derek
B. Leinweber\footnote{E-mail:~derek@phys.washington.edu
{}~$\bullet$~ Telephone: (206)~616--1447 ~$\bullet$~ Fax:
(206)~685--0635 \hfill\break
\null\hspace{6pt}
WWW:~http://www.phys.washington.edu/$\sim$derek/Welcome.html}}

\address{
Department of Physics, Box 351560, University of Washington, Seattle,
WA 98195}

\date{\today}

\maketitle

\begin{abstract}
An examination of the symmetries manifest in the QCD path integral
for current matrix elements reveals various equalities among the
quark sector contributions.  QCD equalities among octet baryon
magnetic moments lead to a determination of the disconnected
sea-quark contribution to nucleon magnetic moments, which is the
most reliable determination in the literature.  Matching QCD
equalities to recent calculations of decuplet baryon magnetic
moments in chiral perturbation theory ($\chi$PT) reveals an
equivalence between $\chi$PT to ${\cal O}(p^2)$ and the simple quark
model with an explicit disconnected sea-quark contribution.  New
insights into SU(3)-flavor symmetry breaking, sea contributions and
constituent quark composition are obtained.  The strangeness
contribution to nucleon magnetic moments is found to be large at
$G_M^s(0) = -0.75 \pm 0.30\ \mu_N$.  The QCD equalities must be
followed by any model which hopes to capture the essence of
nonperturbative QCD.  Not all models are in accord with these
symmetries.
\end{abstract}

\newpage

\narrowtext

\section{Introduction}

The Euclidean path integral formulation of quantum field theory is the
origin of fundamental approaches to the study of quantum
chromodynamics (QCD) in the nonperturbative regime.  While the lattice
regularization of the path integral is the only known approach to {\it
ab initio} determinations of hadron properties, the actual
implementation of the numerical simulations is met with formidable
technical difficulties.  To proceed one must work with unphysically
heavy current quarks and most simulations still involve the quenched
approximation.  Understanding the physics associated with these
approximations is an industry in itself
\cite{cohen92a,leinweber93a,leinweber94d,labrenz94,golterman94}.

   The QCD Sum Rule approach to nonperturbative QCD may also be
formulated in Euclidean space \cite{leinweber95a,leinweber95b}.
However, the Borel improved spectral sum rules are more widely known
and provide better suppression of excited state contaminations.  Here,
one encounters difficulties in determining the vacuum expectation
values of the operators of the operator product expansion (OPE).
Additional uncertainties surround the viability of the continuum
model, utilized to remove excited state contaminations from the OPE.
An in-depth examination of these issues may be found in Ref.\
\cite{leinweber95d}.

In contrast, this investigation focuses on fundamental QCD equalities
based directly on the symmetries of the QCD path integral.  In the
following, it will be clearly stated which properties are fundamental
symmetries of QCD and which properties depend on the actual dynamics
of QCD.  As such, the following equalities must be met by any model
which hopes to capture the essence of nonperturbative QCD.

   The only approximation made in the following discussion of QCD
equalities is the equivalence of the $u$ and $d$ current quark masses.
Hence, we will focus on observables which are not dominated by
isospin violation at the QCD level.  Since the QCD scale parameter
$\Lambda_{\rm QCD} \gg m_d - m_u$, this approximation is generally
accepted to be excellent, and is shared by many approaches probing
hadron structure.

   The outline of this paper is as follows.  Section II reviews the
path integral formalism and the operators used to probe current matrix
elements of low-lying baryons.  Correlation functions are calculated
to reveal the QCD equalities.  Section III discusses the QCD
equalities for octet baryons and demonstrates their utility by
calculating the disconnected sea-quark loop contributions to magnetic
moments.  Section IV addresses the QCD equalities for decuplet baryons
and their relationship to predictions from chiral perturbation theory
($\chi$PT).  Here, quantitative relationships between SU(3) breaking
in the valence and sea sectors are obtained.  Sea contributions are
evaluated and the predictions for baryon magnetic moments from lattice
QCD are updated.  The strangeness contribution to nucleon magnetic
moments is established and compared with other approaches in Section
V.  A comprehensive breakdown of the quark sector contributions to
baryon magnetic moments which may be useful in the development of more
sophisticated models is provided here.  Finally, Section VI reviews
the highlights of QCD equalities.

\section{Formalism}

\subsection{Path Integral}

   The determination of hadron properties in field theoretic
approaches are centered around the QCD vacuum expectation values of
appropriately chosen operators, ${\cal O}_i$.  In the Euclidean path
integral formulation, these vacuum expectation values are given by
\begin{eqnarray}
\lefteqn{
\langle 0 | {\cal O}_1(A_\mu,\psi,\overline \psi)\,
{\cal O}_2(A_\mu,\psi,\overline \psi)\ \cdots  | 0 \rangle = }
\hspace{2cm} \nonumber \\
&&
{1 \over Z}\, \int {\cal D}A_\mu\, {\cal D}\overline \psi\, {\cal
  D}\psi\,
e^{ - S_G(A_\mu) - \overline \psi\, M(A_\mu)\, \psi }\,
\left [ \overline \psi\, \overline \psi \cdots \Gamma(A_\mu)\, \psi\,
  \psi
\cdots \right ] \, ,
\label{pathint}
\end{eqnarray}
with
\begin{equation}
Z =  \int {\cal D}A_\mu\, {\cal D}\overline \psi\, {\cal D}\psi\,
e^{ - S_G(A_\mu) - \overline \psi\, M(A_\mu)\, \psi }\, .
\label{partition}
\end{equation}
Here, $S_G(A_\mu)$ is the gauge action, and $M(A_\mu) = \left (
  \gamma_\mu \, D_\mu + m \right )$ where $D_\mu$ is the covariant
derivative.  A sum over quark flavors with appropriate masses is
implicit.  The explicit dependence of the various terms on the gauge
field $A_\mu$ has been illustrated.  $\Gamma(A_\mu)$ simply signifies
the underlying matrix multiplications in Dirac space and the operator
dependence on $A_\mu$.  The fermion fields are described by a
Grassmann algebra.  The functional integral may be done analytically
to give the well known form
\begin{eqnarray}
\lefteqn{
\langle 0  | {\cal O}_1(A_\mu,\psi,\overline \psi)\,
{\cal O}_2(A_\mu,\psi,\overline \psi)\ \cdots  | 0 \rangle = }
\hspace{2cm} \nonumber \\
& &
{1 \over Z}\, \int {\cal D}A_\mu\,
e^{ - S_G(A_\mu) } \det M(A_\mu) \;
\left [  \Gamma(A_\mu)\, M^{-1}(A_\mu)\, M^{-1}(A_\mu) \cdots \right ]
\,
\label{intpath}
\end{eqnarray}
where each pair of $\psi$ and $\overline \psi$ in the brackets of
(\ref{pathint}) produce $M^{-1}(A_\mu)$, the nonperturbative quark
propagator hereafter referred to as $S$.  The origin of the QCD
equalities lie in the detailed structure of the term in brackets in
(\ref{intpath}).  The remaining factors
\begin{equation}
{\cal D}A_\mu\, e^{ - S_G(A_\mu) }\, \det M(A_\mu)
\end{equation}
are common to all the individual terms in the brackets and more
generally are common to any hadron considered in the approach.  Hence
all the physics differentiating one hadron from the next is contained
in the bracketed term of (\ref{intpath}) which we shall now turn our
attention to in detail.

\subsection{Operators}

   Current matrix elements of hadrons are extracted from the
consideration of a time-ordered product of three operators.
Generally, an operator exciting the hadron of interest from the QCD
vacuum is followed by the current of interest, which in turn is
followed by an operator annihilating the hadron back to the QCD
vacuum.  The operator can take the form
\begin{eqnarray}
\lefteqn{
{\cal O}^{\mu}_{\sigma \tau}
         (t_2, t_1; \vec{p'}, \vec{p}; \Gamma) \,  = } \qquad
\nonumber \\
& & \Gamma^{\beta \alpha} \, T \left ( \int d\vec{x_2} \, e^{-i
\vec{p'} \cdot \vec{x_2} } \, \chi^\alpha_\sigma(x_2) \, \int
d\vec{x_1} e^{+ i \left ( \vec{p'} - \vec{p} \right ) \cdot \vec{x_1}
}\,  j^\mu(x_1) \; \overline \chi^\beta_\tau(0) \right ) \, .
\end{eqnarray}
Here, $\chi$ is a hadron interpolating field usually composed of quark
(and possibly gluon) field operators and Dirac $\gamma$-matrices
designed to isolate the quantum numbers of the hadron under
consideration.  The subscripts $\sigma$ and $\tau$ are optional
Lorentz indices which are required for the excitation of spin-3/2
baryons or vector/axial-vector mesons.  The Lorentz index $\mu$ is
also provided for vector/axial-vector probes of the hadron structure.
Of course, scalar, pseudoscalar, or tensor probes may also be
considered.  The Dirac matrix $\Gamma$ provides for the projection of
various Dirac $\gamma$-structures, and $\alpha$ and $\beta$ are Dirac
indices.  The vectors $\vec{p}$ and $\vec{p'}$ provide for the design
of any momentum transfer.  Euclidean time evolution in $t_1$ and $t_2$
provides a mechanism for the isolation of ground state properties.  To
reveal the QCD equalities, it is sufficient to work in coordinate
space with the operator
\begin{equation}
{\cal O}^{\mu}_{\sigma \tau}(x_2, x_1) =
T \left ( \chi^\alpha_\sigma(x_2) \, j^\mu(x_1) \, \overline
\chi^\beta_\tau(0) \right ) \, .
\end{equation}

\subsection{Interpolating Fields}

   A gauge invariant operator having maximal overlap with the ground
state hadron is local and of minimal dimension.  These criteria
alone are sufficient to uniquely define the $\Delta$-baryon
interpolating field.  For $\Delta^{++}$ the form is
\begin{equation}
\chi_\mu^{\Delta^{++}}(x) =
\epsilon^{abc} \left ( u^{Ta}(x)\, C \gamma_\mu\,
                                u^b(x) \right ) u^c(x).
\label{chiDelta}
\end{equation}
Here $u$ denotes the up-quark field, and the indices $a$, $b$ and $c$
are color indices.  The antisymmetric tensor $\epsilon^{a b c}$ places
the three quarks in a color singlet state.  $C$ is the charge
conjugation operator, and $T$ denotes transpose.  The interpolators
coupling to other charge states of $\Delta$ may be obtained from
equation (\ref{chiDelta}) by using the isospin-three-component
lowering operator.  In particular
\begin{equation}
\chi_\mu^{\Delta^{+}}(x) =
{1 \over \sqrt{3} } \; \epsilon^{abc} \left [
2 \left ( u^{Ta}(x)\, C \gamma_\mu\, d^b(x) \right )
u^c(x)
+ \left ( u^{Ta}(x)\, C \gamma_\mu\, u^b(x) \right )
d^c(x)\, \right ] \, .
\label{deltapif}
\end{equation}
Other decuplet baryon interpolating fields are obtained with the
appropriate substitutions of $u(x),\ d(x)\ \to\ u(x),\ d(x)$ or
$s(x)$.

   The uniqueness of the $\Delta^{++}$ interpolator is easily
demonstrated \cite{ioffe81} by considering the general local current
with $I = 3/2$ and $J = 3/2$.  Consider
\begin{equation}
\chi_\mu^{\Delta^{++}} = \epsilon^{abc} \left ( u^{Ta}\, C\,
\Gamma_\epsilon\, u^b \right ) \gamma_\mu\, \Gamma^{\epsilon'}\,
u^c \, ,
\label{genDelta}
\end{equation}
where $\Gamma_\epsilon$ may take on any of the sixteen Dirac
$\gamma$-matrices.  By transposing the scalar quantity in parentheses
one finds
\begin{equation}
\epsilon^{abc} u^{Ta}\, C \, u^b =
  \epsilon^{abc} u^{Ta}\, C \gamma_5\, u^b =
  \epsilon^{abc} u^{Ta}\, C \gamma_\mu \gamma_5\, u^b = 0.
\end{equation}
An application of the Fierz transformations reveals
\begin{equation}
\epsilon^{abc} \bigl ( u^{Ta}\, C \sigma_{\rho \lambda}\, u^b \bigr )
\gamma_\mu \sigma^{\rho \lambda}\, u^c = - {1\over 2}
\epsilon^{abc} \bigl ( u^{Ta}\, C \sigma_{\rho \lambda}\, u^b \bigr )
\gamma_\mu \sigma^{\rho \lambda}\, u^c = 0.
\end{equation}
Another application of the Fierz transformations provides the relation
\begin{equation}
\epsilon^{abc} \bigl ( u^{Ta}\, C \gamma_\lambda\, u^b \bigr )
\sigma^{\lambda \mu}\, u^c =
i \epsilon^{abc} \bigl ( u^{Ta}\, C \gamma_\mu\, u^b \bigr ) u^c ,
\end{equation}
equating the only non-zero variants of (\ref{genDelta}).

   It is well established that there are two local interpolating
fields of minimal dimension for the nucleon.  In lattice calculations,
the commonly used interpolating field for the proton has the form
\begin{equation}
\chi_1^p(x) = \epsilon^{abc}
                 \left ( u^{Ta}(x) C \gamma_5 d^b(x) \right ) u^c(x)
\, .
\label{chiN1}
\end{equation}
In the QCD Sum Rule approach, it is common to find linear combinations
of this interpolating field and
\begin{equation}
\chi_2^p(x) = \epsilon^{abc}
                 \left ( u^{Ta}(x) C d^b(x) \right ) \gamma_5 u^c(x)
\, ,
\label{chiN2}
\end{equation}
which vanishes in the nonrelativistic limit.  Interpolating fields for
the other members of the baryon octet containing a doubly represented
quark flavor may be obtained from (\ref{chiN1}) and (\ref{chiN2}) with
the appropriate substitutions of quark field operators.

\subsection{Correlation Functions}

   To begin, consider the two-point function for $\Delta^+$.  Using
(\ref{chiDelta}) and contracting out pairs of quark field operators in
accord with (\ref{intpath}) one has
\begin{eqnarray}
\lefteqn{
T \left ( \chi_\mu^{\Delta^+}(x) \,
\overline \chi_\nu^{\Delta^+}(0) \right )  =
{1 \over 3} \; \epsilon^{abc} \epsilon^{a'b'c'}  \biggl \{ }
\nonumber \\
&& \;\;\: 4 S_u^{a a'} \,
                       \gamma_\nu \, C S_u^{T b b'} C \, \gamma_\mu \,
S_d^{c c'}
   + 4 S_u^{a a'}\, \gamma_\nu \, C S_d^{T b b'} C \, \gamma_\mu \,
S_u^{c c'}
   + 4 S_d^{a a'}\, \gamma_\nu \, C S_u^{T b b'} C \, \gamma_\mu \,
S_u^{c c'} \nonumber \\
&& + 2 S_u^{a a'} \, tr \left [ \gamma_\nu \,
  C S_u^{T b b'} C \, \gamma_\mu \, S_d^{c c'} \right ]
   + 2 S_u^{a a'} \, tr \left [ \gamma_\nu \,
  C S_d^{T b b'} C \, \gamma_\mu \, S_u^{c c'} \right ]
   + 2 S_d^{a a'} \, tr \left [ \gamma_\nu \,
  C S_u^{T b b'} C \, \gamma_\mu \, S_u^{c c'} \right ] \biggr \}
\nonumber \\
\label{DeltaTwoPt}
\end{eqnarray}
\narrowtext
where the quark-propagator $S_u^{a a'} = T \left ( u^a(x), \overline
u^{a'}(0) \right )$ and similarly for other quark flavors.
$SU(3)$-flavor symmetry is obviously displayed in this equation.

\begin{figure}[t]
\begin{center}
\epsfxsize=8truecm
\leavevmode
\epsfbox{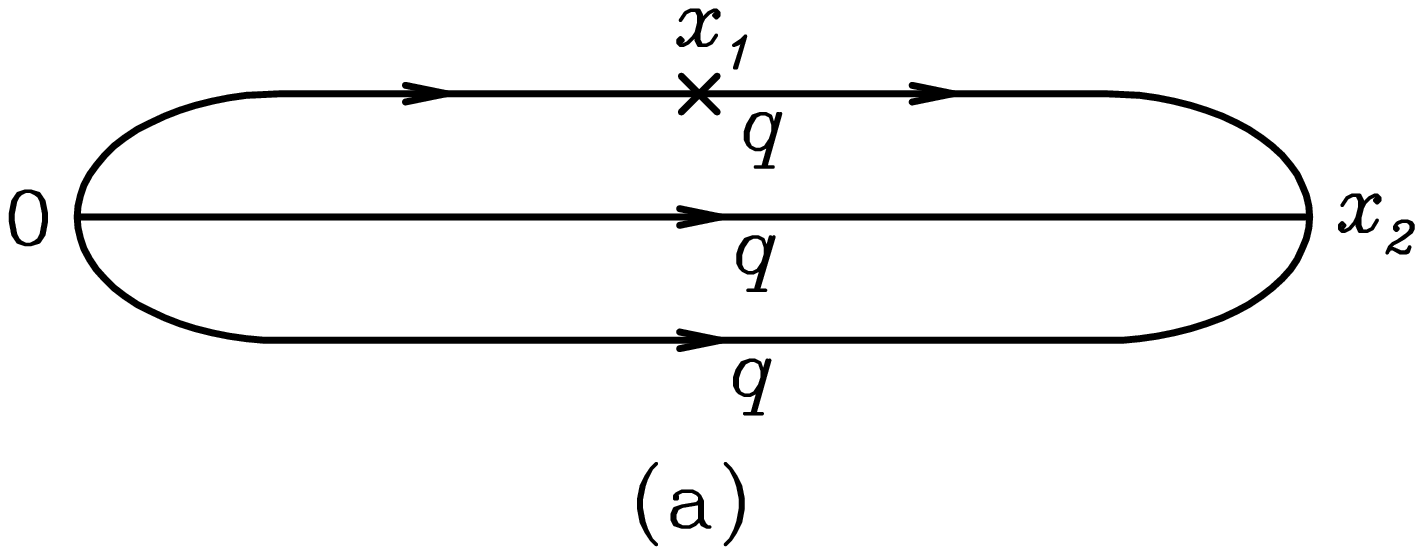}
\end{center}
\vspace{12pt}
\begin{center}
\epsfxsize=8truecm
\leavevmode
\epsfbox{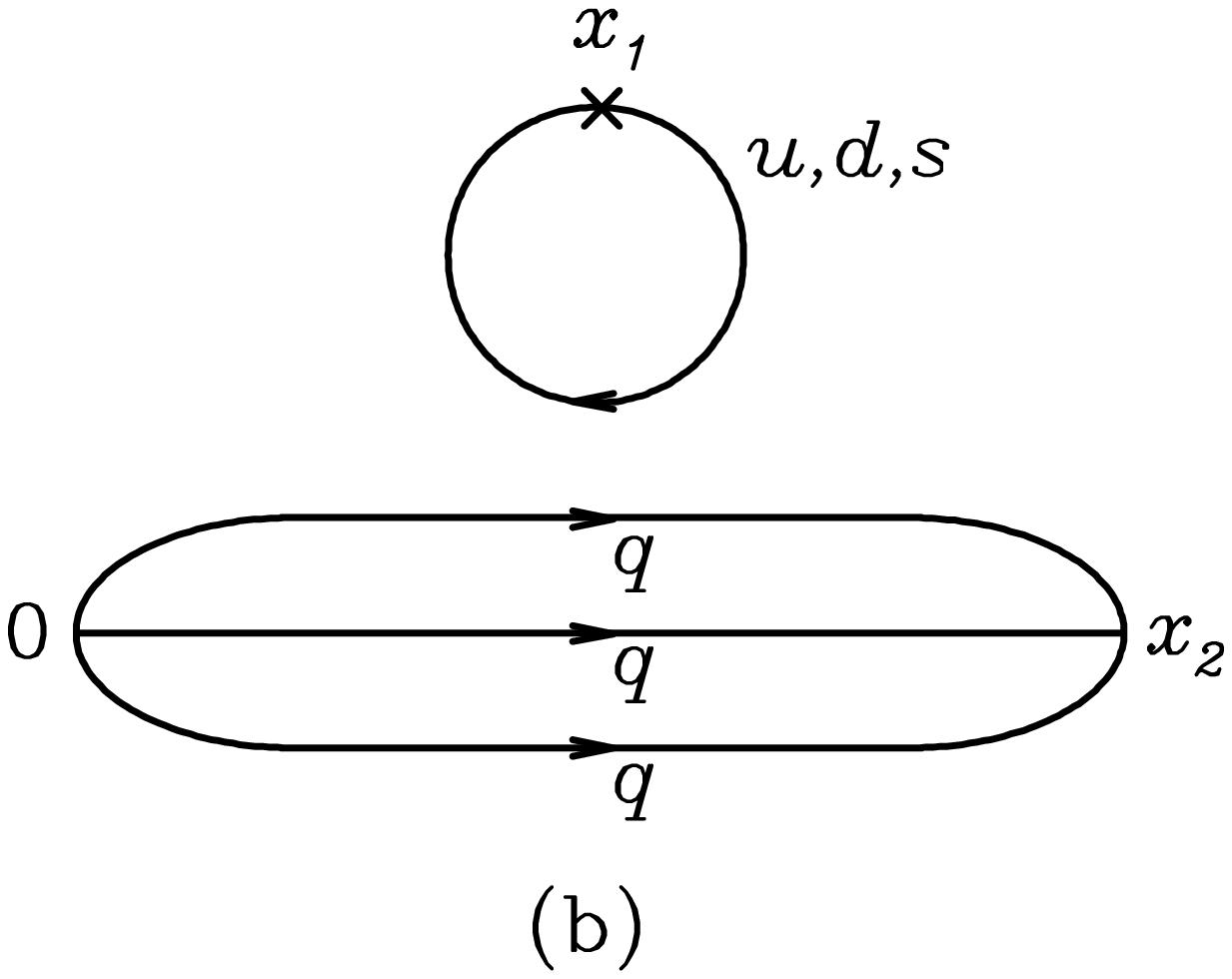}
\end{center}
\caption{Diagrams illustrating the two topologically different
insertions of the current indicated by $\times$.  These skeleton
diagrams for the connected (a) and disconnected (b) current insertions
may be dressed by an arbitrary number of gluons.}
\label{topology}
\end{figure}

   In determining the three point function, one encounters two
topologically different ways of performing the current insertion.
Figure \ref{topology} displays skeleton diagrams for these two
insertions.  These diagrams may be dressed with an arbitrary number of
gluons.  Diagram (a) illustrates the connected insertion of the
current to one of the valence\footnote{It should be noted that the
term ``valence'' used here differs with that commonly used
surrounding discussions of deep-inelastic structure functions.  Here
``valence'' simply describes the quark whose quark flow line runs
continuously from $0 \to x_2$.  These lines can flow backwards as well
as forwards in time and therefore have a sea contribution associated
with them \protect\cite{cohen92a}.}
quarks of the baryon.  It is here that the Pauli-blocking in the sea
contributions are taken into account.  Diagram (b) accounts for the
alternative time ordering where the current first produces a
disconnected $q \, \overline q$ pair which in turn interacts with the
valence quarks of the baryon via gluons.

   The number of terms in the three-point function is four times that
in (\ref{DeltaTwoPt}).  The correlation function relevant to the
$\Delta^+$ current matrix element is
\begin{eqnarray}
\lefteqn{
T \left ( \chi_\mu^{\Delta^+}(x_2) \, j^\mu(x_1) \,
\overline \chi_\nu^{\Delta^+}(0) \right )  =
{1 \over 3} \; \epsilon^{abc} \epsilon^{a'b'c'}  \biggl \{ }
\nonumber \\
&& \;\;\: 4 \widehat S_u^{a a'} \, \gamma_\nu \, C S_u^{T b b'} C \,
\gamma_\mu \, S_d^{c c'}
+ 4 \widehat S_u^{a a'}\, \gamma_\nu \, C S_d^{T b b'} C \,
\gamma_\mu \, S_u^{c c'}
+ 4 \widehat S_d^{a a'}\, \gamma_\nu \, C S_u^{T b b'} C \,
\gamma_\mu \, S_u^{c c'} \nonumber \\
&& + 4 S_u^{a a'} \, \gamma_\nu \, C \widehat S_u^{T b b'} C \,
\gamma_\mu \, S_d^{c c'}
   + 4 S_d^{a a'}\, \gamma_\nu \, C \widehat S_u^{T b b'} C \,
\gamma_\mu \, S_u^{c c'}
   + 4 S_u^{a a'}\, \gamma_\nu \, C \widehat S_d^{T b b'} C \,
\gamma_\mu \, S_u^{c c'} \nonumber \\
&& + 4 S_u^{a a'}\, \gamma_\nu \, C S_d^{T b b'} C \, \gamma_\mu \,
\widehat S_u^{c c'}
   + 4 S_d^{a a'}\, \gamma_\nu \, C S_u^{T b b'} C \, \gamma_\mu \,
\widehat S_u^{c c'}
   + 4 S_u^{a a'} \, \gamma_\nu \, C S_u^{T b b'} C \,
\gamma_\mu \, \widehat S_d^{c c'} \nonumber \\
&& + 2 \widehat S_u^{a a'} \, tr \left [ \gamma_\nu \,
  C S_u^{T b b'} C \, \gamma_\mu \, S_d^{c c'} \right ]
   + 2 \widehat S_u^{a a'} \, tr \left [ \gamma_\nu \,
  C S_d^{T b b'} C \, \gamma_\mu \, S_u^{c c'} \right ]
   + 2 \widehat S_d^{a a'} \, tr \left [ \gamma_\nu \,
  C S_u^{T b b'} C \, \gamma_\mu \, S_u^{c c'} \right ]
\nonumber \\
&& + 2 S_u^{a a'} \, tr \left [ \gamma_\nu \,
  C \widehat S_u^{T b b'} C \, \gamma_\mu \, S_d^{c c'} \right ]
   + 2 S_d^{a a'} \, tr \left [ \gamma_\nu \,
  C \widehat S_u^{T b b'} C \, \gamma_\mu \, S_u^{c c'} \right ]
   + 2 S_u^{a a'} \, tr \left [ \gamma_\nu \,
  C \widehat S_d^{T b b'} C \, \gamma_\mu \, S_u^{c c'} \right ]
\nonumber \\
&& + 2 S_u^{a a'} \, tr \left [ \gamma_\nu \,
  C S_d^{T b b'} C \, \gamma_\mu \, \widehat S_u^{c c'} \right ]
   + 2 S_d^{a a'} \, tr \left [ \gamma_\nu \,
  C S_u^{T b b'} C \, \gamma_\mu \, \widehat S_u^{c c'} \right ]
   +  2 S_u^{a a'} \, tr \left [ \gamma_\nu \,
  C S_u^{T b b'} C \, \gamma_\mu \, \widehat S_d^{c c'} \right ]
\biggr \}
\nonumber \\
&& + \sum_{q = u,\, d,\, s} e_q\, \sum_i
tr \left [ S_q^{ii}(x_1,x_1) \, \gamma_\mu \right ]
{1 \over 3} \; \epsilon^{abc} \epsilon^{a'b'c'}  \biggl \{
\nonumber \\
&& \;\;\: 4 S_u^{a a'} \,
                       \gamma_\nu \, C S_u^{T b b'} C \, \gamma_\mu \,
S_d^{c c'}
  + 4 S_u^{a a'}\, \gamma_\nu \, C S_d^{T b b'} C \, \gamma_\mu \,
S_u^{c c'}
  + 4 S_d^{a a'}\, \gamma_\nu \, C S_u^{T b b'} C \, \gamma_\mu \,
S_u^{c c'} \nonumber \\
&& + 2 S_u^{a a'} \, tr \left [ \gamma_\nu \,
  C S_u^{T b b'} C \, \gamma_\mu \, S_d^{c c'} \right ]
   + 2 S_u^{a a'} \, tr \left [ \gamma_\nu \,
  C S_d^{T b b'} C \, \gamma_\mu \, S_u^{c c'} \right ]
   +  2 S_d^{a a'} \, tr \left [ \gamma_\nu \,
  C S_u^{T b b'} C \, \gamma_\mu \, S_u^{c c'} \right ]
\biggr \}
\nonumber \\
\label{Delta3pt}
\end{eqnarray}
where
\begin{equation}
\widehat S_q^{aa'}(x_2,x_1,0) = e_q\, \sum_{i} S_q^{ai}(x_2,x_1) \,
\gamma_\mu \, S_q^{i a'}(x_1,0) \, ,
\end{equation}
denotes the connected insertion of the probing current to a quark of
charge $e_q$.  Here we have explicitly selected the electromagnetic
current.  However, the following discussion may be generalized to any
quark-field-based current operator bilinear in the quark fields.

   The latter term of (\ref{Delta3pt}) accounts for the loop
contribution depicted in figure \ref{topology}b.  The sum over the
quarks running around the loop has been restricted to the flavors
relevant to the ground state baryon octet and decuplet.  In the
SU(3)-flavor limit the sum vanishes for the electromagnetic current.
However, the heavier strange quark mass allows for a nontrivial
result.  Due to the technical difficulties of numerically estimating
$M^{-1}$ for the squared lattice volume of diagonal spatial indices at
$q^2 \ne 0$, these contributions have been omitted from previous
lattice calculations of electromagnetic structure.  For other
observables such as the scalar density or forward matrix elements of
the axial vector current relevant to the spin of the baryon, the
``charges'' running around the loop do not sum to zero.  In this case
the second term of (\ref{Delta3pt}) can be just as significant as the
connected term \cite{kuramashi93,dong94}.

   An examination of (\ref{Delta3pt}) reveals complete symmetry among
the quark flavors in the correlation function.  For example, wherever
a $d$ quark appears in the correlator, a $u$ quark also appears in the
same position in another term.  An interesting consequence of this is
that the connected insertion of the electromagnetic current for
$\Delta^0$ vanishes.  All electromagnetic properties of the $\Delta^0$
have their origin strictly in the disconnected loop contribution.
Physically, what this means is that the valence wave function for each
of the quarks in the $\Delta$ resonances are identical.

   To further illustrate this symmetry, we turn to another system in
which this symmetry is broken; namely the nucleon.  The correlation
function relevant to proton matrix elements obtained from
(\ref{chiN1}) is
\begin{eqnarray}
\lefteqn{
T \left ( \chi_1^{p}(x_2) \, j^\mu(x_1) \,
\overline \chi_1^{p}(0) \right )  =
\epsilon^{abc} \epsilon^{a'b'c'}  } \qquad\qquad
\nonumber \\
\biggl \{
&& \;\;\:
\widehat S_u^{a a'} \, {\rm tr} \left ( S_u^{b b'} \, \gamma_5 C \,
  S_d^{T
c c'} C^{-1} \gamma_5 \right )
+ \widehat S_u^{a a'} \, \gamma_5 C \, S_d^{T b b'} C^{-1} \gamma_5 \,
S_u^{c c'} \nonumber \\
&& + S_u^{a a'} \, {\rm tr} \left ( \widehat S_u^{b b'} \, \gamma_5 C
\, S_d^{T c c'} C^{-1} \gamma_5 \right )
+ S_u^{a a'} \, \gamma_5 C \, S_d^{T b b'} C^{-1} \gamma_5 \, \widehat
S_u^{c c'} \nonumber \\
&& + S_u^{a a'} \, {\rm tr} \left ( S_u^{b b'} \, \gamma_5 C \,
\widehat S_d^{T c c'} C^{-1} \gamma_5 \right )
+ S_u^{a a'} \, \gamma_5 C \, \widehat S_d^{T b b'} C^{-1} \gamma_5 \,
S_u^{c c'}
\biggr \} \nonumber \\
&& + \sum_{q = u,\, d,\, s} e_q\, \sum_i
tr \left [ S_q^{ii}(x_1,x_1) \, \gamma_\mu \right ]
\epsilon^{abc} \epsilon^{a'b'c'}  \nonumber \\
&&\quad \biggl \{
S_u^{a a'} \, {\rm tr} \left ( S_u^{b b'} \, \gamma_5 C \, S_d^{T
c c'} C^{-1} \gamma_5 \right )
+ S_u^{a a'} \, \gamma_5 C \, S_d^{T b b'} C^{-1} \gamma_5 \, S_u^{c
  c'}
\biggr \} \, ,
\label{Nucl3pt1}
\end{eqnarray}
Here we see very different roles played by $u$ and $d$ quarks in the
correlation function.  The neutron correlation function is obtained
with the exchange of $u \leftrightarrow d$ in (\ref{Nucl3pt1}).  The
absence of equivalence for $u$ and $d$ contributions allows the
connected quark sector to give rise to a nontrivial neutron charge
radius, a large neutron magnetic moment, or a violation of the
Gottfried sum rule.  Perhaps it is also worth noting that the relative
contributions of $u$ and $d$ quarks to nucleon moments depends on the
dynamics of QCD.  Numerical simulations indicate that the relative
contributions of the $u$ and $d$ valance quarks to nucleon magnetic
moments are very different from the SU(6)-spin-flavor symmetric wave
functions of the simple quark model \cite{leinweber91}.

   Another interesting point to emphasize, is that there is no simple
relationship between the properties of a particular quark flavor bound
in different baryons.  For example, the correlator for $\Sigma^+$ is
given by (\ref{Nucl3pt1}) with $d \to s$.  Hence, a $u$-quark
propagator in $\Sigma^+$ is multiplied by an $s$-quark propagator,
whereas in the proton the $u$-quark propagators are multiplied by a
$d$-quark propagator.  The different mass of the neighboring quark
gives rise to an environment sensitivity in the $u$-quark
contributions to
observables\cite{leinweber91,leinweber92a,leinweber92b,leinweber93e,%
leinweber93b,leinweber94f,leinweber95i}.  This point sharply
contrasts the concept of an intrinsic quark property which is
independent of the quark's environment.  This concept of an intrinsic
quark property is a fundamental foundation of many constituent based
quark models and is not in accord with QCD.

   The symmetry of the correlation function obtained from $\chi_2^p$
of (\ref{chiN2}) is identical to that of (\ref{Nucl3pt1}).  The
relevant correlation function is
\begin{eqnarray}
\lefteqn{
T \left ( \chi_2^{p}(x_2) \, j^\mu(x_1) \,
\overline \chi_2^{p}(0) \right )  =
\epsilon^{abc} \epsilon^{a'b'c'}  } \qquad\qquad
\nonumber \\
\biggl \{
&& \;\;\:
\gamma_5 \, \widehat S_u^{a a'} \, \gamma_5 \, {\rm tr} \left ( S_u^{b
    b'} \, C \, S_d^{T
c c'} C^{-1} \right )
+ \gamma_5 \, \widehat S_u^{a a'} \,  C \, S_d^{T b b'} C^{-1}  \,
S_u^{c c'} \gamma_5 \nonumber \\
&& + \gamma_5 \, S_u^{a a'} \, \gamma_5 \, {\rm tr} \left ( \widehat
  S_u^{b b'} \,  C
\, S_d^{T c c'} C^{-1} \right )
+ \gamma_5 \, S_u^{a a'} \, C \, S_d^{T b b'} C^{-1} \, \widehat
S_u^{c c'} \gamma_5 \nonumber \\
&& + \gamma_5 \, S_u^{a a'} \, \gamma_5 \, {\rm tr} \left ( S_u^{b b'}
  \, C \,
\widehat S_d^{T c c'} C^{-1} \right )
+ \gamma_5 \, S_u^{a a'} \, C \, \widehat S_d^{T b b'} C^{-1}  \,
S_u^{c c'} \gamma_5
\biggr \} \nonumber \\
&& + \sum_{q = u,\, d,\, s} e_q\, \sum_i
tr \left [ S_q^{ii}(x_1,x_1) \, \gamma_\mu \right ]
\epsilon^{abc} \epsilon^{a'b'c'}  \nonumber \\
&&\quad \biggl \{
\gamma_5 \, S_u^{a a'} \, \gamma_5 \, {\rm tr} \left ( S_u^{b b'} \,
  C \, S_d^{T
c c'} C^{-1} \right )
+ \gamma_5 \, S_u^{a a'} \, C \, S_d^{T b b'} C^{-1} \, S_u^{c c'}
\gamma_5
\biggr \} \, .
\label{Nucl3pt2}
\end{eqnarray}
The interference of $\chi_1^p$ and  $\chi_2^p$ provide a similar
symmetry.

   As a final point, it is important to note that the symmetry of $u
\leftrightarrow d$ for describing the current matrix elements of the
neutron in terms of the proton is always satisfied when the
disconnected loop contribution is separated from the valence sector as
in (\ref{Nucl3pt1}) or (\ref{Nucl3pt2}).  Many models fail to include
the loop contribution explicitly, but rather break isospin symmetry
between the $u$ and $d$ quarks by absorbing the loop contribution into
the definition of the constituent quark.  This leads to the common
misinterpretation that isospin symmetry breaking in octet baryons is
large.  When the loop contribution is absorbed into the valence quark
contribution, an exchange of $u$ and $d$ contributions does not
necessarily correctly describe the neutron in terms of the proton, or
vice-versa.

   It is important to estimate the size of such disconnected loop
contributions in the nucleon.  As we shall see, an estimate of the
strangeness contribution to nucleon matrix elements can be obtained
{}from the following QCD equalities.

\section{Octet Baryon Symmetries and Loop Contributions}
\label{OctetSym}

   Equations (\ref{Nucl3pt1}) and (\ref{Nucl3pt2}) provide the
following equalities for current matrix elements of octet baryons.
\begin{mathletters}
\begin{eqnarray}
p &=& e_u\, D_N + e_d\, S_N + O_N  \, ,  \\
n &=& e_d\, D_N + e_u\, S_N + O_N  \, ,  \\
\Sigma^+ &=& e_u\, D_\Sigma + e_s\, S_\Sigma + O_\Sigma  \, ,  \\
\Sigma^- &=& e_d\, D_\Sigma + e_s\, S_\Sigma + O_\Sigma  \, ,  \\
\Xi^0 &=& e_s\, D_\Xi + e_u\, S_\Xi + O_\Xi  \, ,  \\
\Xi^- &=& e_s\, D_\Xi + e_d\, S_\Xi + O_\Xi  \, ,%
\end{eqnarray}%
\end{mathletters}%
Here, $D$, $S$, and $O$ represent contributions from the doubly
represented valence quark flavor, the singly represented valence
flavor, and the disconnected loop sector respectively.  Subscripts
allow for environment sensitivity, $e$ indicates the quark flavor
``charge'' and is not restricted to electric charge, and the baryon
label represents the observable corresponding to the charge, momentum
transfer and Lorentz component(s) of the probing current.  These are
QCD equalities and must be reproduced by any model which hopes to
reflect the properties of QCD when the light current quark masses are
isospin symmetric.

   There are two important features here which are often neglected in
model formulations.  Isospin symmetric models based on the valence
sector omit the disconnected sea-quark loop contribution, $O$.  Often,
no provision is made for the environment sensitivity of the valence
sector contributions.

   On the surface, it appears we have described six quantities in
terms of nine parameters.  However, the QCD equalities are more
generally applicable.  They are not restricted to the bulk baryon
properties, but also provide information on the individual quark
sector contributions, to be measured at CEBAF.  The equalities define
a pattern for quark sector contributions which may be useful in the
development of more sophisticated models.

   Focusing now on electromagnetic properties, the disconnected loop
contributions may be isolated in the following favorable forms,
\begin{mathletters}
\begin{eqnarray}
O_N &=& {1 \over 3} \left \{ 2\, p + n - {D_N \over D_\Sigma} \left (
\Sigma^+ - \Sigma^- \right ) \right \} \, \label{disconnN1} \\
O_N &=& {1 \over 3} \left \{ p + 2\, n - {S_N \over S_\Xi} \left (
\Xi^0 - \Xi^- \right ) \right \} \, \label{disconnN2} \\
O_\Sigma &=& {1 \over 3} \left \{ \Sigma^+ + 2\, \Sigma^- + {S_\Sigma
\over S_\Xi} \left ( \Xi^0 - \Xi^- \right ) \right \} \,
\label{disconnSigma} \\
O_N &=& {1 \over 3} \left \{ \Xi^0 + 2\, \Xi^- + {D_\Xi \over
D_\Sigma} \left ( \Sigma^+ - \Sigma^- \right ) \right \} \,
\label{disconnXi}%
\end{eqnarray}%
\label{disconn}%
\end{mathletters}%
The ratios of doubly or singly represented quark contributions
appearing in (\ref{disconn}) account for the environment sensitivity
of the quark sector contribution.  In terms of quark flavors, the
ratios appearing in (\ref{disconnN1}) and (\ref{disconnN2}) for the
nucleon are
\begin{equation}
{D_N \over D_\Sigma} = {u_p \over u_\Sigma} = {d_n \over d_\Sigma}\, ,
\quad {\rm and} \quad
{S_N \over S_\Xi} = {d_p \over d_\Xi} = {u_n \over u_\Xi} \, .
\end{equation}
In many quark models, these ratios are simply taken to be one.  Hence,
by identifying the loop contributions in which the leading terms of
(\ref{disconn}) dominate the total contribution, a determination of
the disconnected sea contribution may be obtained with minimal model
dependence.

   For magnetic moments, $O \equiv \mu_l$, and equation
(\ref{disconn}) takes the form
\begin{mathletters}
\begin{eqnarray}
\mu_l^N &=& {1 \over 3} \left \{ 3.673 - {\mu_u^p \over \mu_u^\Sigma}
\left ( 3.618 \right ) \right \} \, , \label{mu_l^N1} \\
\mu_l^N &=& {1 \over 3} \left \{ -1.033 - {\mu_u^n \over \mu_u^\Xi}
\left ( -0.599 \right ) \right \} \, , \label{mu_l^N2} \\
\mu_l^\Sigma &=& {1 \over 3} \left \{ 0.138 + {\mu_s^\Sigma \over
\mu_d^\Xi} \left ( -0.599 \right ) \right \} \, , \label{mu_l^S} \\
\mu_l^\Xi &=& {1 \over 3} \left \{ -2.551 + {\mu_s^\Xi \over
\mu_d^\Sigma} \left ( 3.618 \right ) \right \} \, , \label{mu_l^X}
\end{eqnarray}%
\label{magnetic}%
\end{mathletters}%
where the moments are in units of nuclear magnetons $(\mu_N)$.
Equation (\ref{mu_l^N2}) provides a favorable case for a determination
of the disconnected sea contribution to the nucleon's magnetic moment
with minimal model dependence.  Taking the simple quark model ratio of
$\mu_u^n / \mu_u^\Xi = 1$ provides $\mu_l^N = -0.14\ \mu_N$.

   To improve on this estimate, we turn to the lattice QCD
calculations of environment sensitivity for these moments
\cite{leinweber91,leinweber92b}.  Because of the nature of the ratios
involved, the systematic uncertainties in the lattice QCD calculations
are expected to be small relative to the statistical uncertainties.
Statistical uncertainties for the relevant ratios are estimated via a
third-order single-elimination jackknife
\cite{leinweber91,leinweber92b}.  The following ratios of magnetic
moments are found
\begin{equation}
{D_N \over D_\Sigma}   = 1.136 \pm 0.078 \, , \quad
{S_N \over S_\Xi}      = 0.721 \pm 0.457 \, , \quad
{S_\Sigma \over S_\Xi} = 0.390 \pm 0.244 \, , \quad
{D_\Xi \over D_\Sigma} = 0.585 \pm 0.039 \, .
\label{MagMomRat}
\end{equation}
The latter two ratios are to be compared with 0.65 in the simple quark
model \cite{rpp94}.  A combination of these uncertainties with the
experimental uncertainties in quadrature, provides the following
predictions, in units of $\mu_N$, corresponding to (\ref{mu_l^N1})
through (\ref{mu_l^X})
\begin{equation}
\mu_l^N      = -0.15 \pm 0.09 \, , \quad
\mu_l^N      = -0.20 \pm 0.09 \, , \quad
\mu_l^\Sigma = -0.03 \pm 0.05 \, , \quad
\mu_l^\Xi    = -0.14 \pm 0.05 \, . \quad
\label{OctetLoops}
\end{equation}
Despite very different uncertainties in the quark sector ratios, both
(\ref{mu_l^N1}) and (\ref{mu_l^N2}) yield similar estimates for the
disconnected sea contribution to the nucleon's magnetic moment.  A
weighted average provides
\begin{equation}
\mu_l^N      = -0.17 \pm 0.07\ \mu_N \, .
\end{equation}
In view of the minimal model dependence associated with
(\ref{mu_l^N2}) this estimate is the most reliable estimate for the
disconnected loop contribution to the nucleon's magnetic moment to
date.  These results for the nucleon agree with another estimate
obtained through an examination of experimental violations of the
Sachs sum rule for magnetic moments \cite{leinweber92a}. There
$\mu_l^N = -0.19 \pm 0.09\ \mu_N$.  The disconnected sea contribution
to the nucleon moment is the order of 10\%.  In light of the precision
data, such a contribution is relatively significant.

   The results for $\mu_l^\Sigma$ and $\mu_l^\Xi$ suggest
contrasting views for the environment sensitivity of the disconnected
loop contributions.  However, the results are similar at
the one standard deviation level.  The relation of (\ref{mu_l^S}) for
$\mu_l^\Sigma$ is particularly unfavorable and the result depends more
sensitively on the ratio of quark flavor contributions.

\section{Decuplet Baryon Symmetries and Loop Contributions}

\subsection{QCD Equalities and Constituent Quark Composition}

   The maximal symmetry of the decuplet baryons, and the four charge
states of the $\Delta$ provide a much richer environment for the
foundation of useful QCD equalities.  The symmetries of
(\ref{Delta3pt}) provide the following relationships among current
matrix elements of decuplet baryons
\begin{mathletters}
\begin{eqnarray}
\Delta^{++} &=& \left ( 3\, e_u + 0\, e_d \right )\, L_\Delta
                + O_\Delta  \, ,  \\
\Delta^+    &=& \left ( 2\, e_u + 1\, e_d \right )\, L_\Delta
                + O_\Delta  \, ,  \\
\Delta^0    &=& \left ( 1\, e_u + 2\, e_d \right )\, L_\Delta
                + O_\Delta  \, ,  \\
\Delta^-    &=& \left ( 0\, e_u + 3\, e_d \right )\, L_\Delta
                + O_\Delta  \, ,  \\
\Sigma^{*+} &=& 2 \, e_u \, L_{\Sigma^*}  + e_s \, H_{\Sigma^*}
                + O_{\Sigma^*} \, , \\
\Sigma^{*0} &=& \left (e_u + e_d \right )\, L_{\Sigma^*} + e_s
H_{\Sigma^*}
                + O_{\Sigma^*} \, , \\
\Sigma^{*-} &=& 2 \, e_d \, L_{\Sigma^*}  + e_s \, H_{\Sigma^*}
                + O_{\Sigma^*} \, , \\
\Xi^{*0}    &=& 2 \, e_s \, H_{\Xi^*} + e_u \, L_{\Xi^*} + O_{\Xi^*}
\, ,  \\
\Xi^{*-}    &=& 2 \, e_s \, H_{\Xi^*} + e_d \, L_{\Xi^*} + O_{\Xi^*}
\, , \\
\Omega^-    &=& 3 \, e_s \, H_\Omega                     +
O_\Omega \, .
\end{eqnarray}%
\end{mathletters}%
Here $L$, $H$ and $O$ denote light, heavy (strange) and disconnected
sea-quark loops respectively.  The subscript allows for environment
sensitivity.  As before, the charge factors are not necessarily
restricted to electromagnetic charge.

   Specializing to electromagnetic properties, the symmetries are
\begin{mathletters}
\begin{eqnarray}
\Delta^{++} &=& +2 \, L_\Delta + O_\Delta  \, ,  \\
\Delta^+    &=& +1 \, L_\Delta + O_\Delta  \, ,  \\
\Delta^0    &=& +0 \, L_\Delta + O_\Delta  \, ,  \\
\Delta^-    &=& -1 \, L_\Delta + O_\Delta  \, ,  \\
\Sigma^{*+} &=& +{4 \over 3} \, L_{\Sigma^*} - {1 \over 3}\,
H_{\Sigma^*}
                + O_{\Sigma^*} \, , \\
\Sigma^{*0} &=& +{1 \over 3} \, L_{\Sigma^*} - {1 \over 3}\,
H_{\Sigma^*}
                + O_{\Sigma^*} \, , \\
\Sigma^{*-} &=& -{2 \over 3} \, L_{\Sigma^*} - {1 \over 3}\,
H_{\Sigma^*}
                + O_{\Sigma^*} \, , \\
\Xi^{*0}    &=& -{2 \over 3} \, H_{\Xi^*} + {2 \over 3}\, L_{\Xi^*}
                + O_{\Xi^*} \, ,  \\
\Xi^{*-}    &=& -{2 \over 3} \, H_{\Xi^*} - {1 \over 3}\, L_{\Xi^*}
                + O_{\Xi^*} \, , \\
\Omega^-    &=& - H_\Omega + O_\Omega \, .
\end{eqnarray}%
\label{decupletEq}%
\end{mathletters}%
The $\Delta^0$ properties are a direct reflection of SU(3) flavor
symmetry breaking in the disconnected sea.  The central point here is
that the disconnected sea-quark loop contribution cannot be absorbed
into the connected contribution.  This observation provides some
rather interesting insight into the composition of a constituent
quark.

   In the simple constituent quark model, decuplet baryon magnetic
moments are obtained by summing the intrinsic moments of the
constituent quarks.  In this case, $L$ and $H$ of (\ref{decupletEq})
are assigned global values independent of the baryon, and the
disconnected loop contribution, $O$, vanishes.  Thus the constituent
quark is void of any disconnected sea-quark loop physics.  It is
composed of a current quark dressed with nonperturbative glue, and as
such, has a sea-quark component associated with the $Z$-graphs of the
current quark.  This discussion carries over to octet baryons provided
the constituent quark properties are isospin symmetric, as in $\mu_u =
- 2\, \mu_d$, etc.

In the simplest constituent quark model based on SU(6) symmetry, the
proton moment is given by
\begin{equation}
\mu_p = {4 \over 3} \, \mu_u - {1 \over 3} \, \mu_d + \mu_l \, ,
\end{equation}
with $\mu_u = - 2\, \mu_d$ and the disconnected loop contribution,
$\mu_l = 0$.  However, since the factors 4/3 and $-1/3$ sum to one, it
is possible to absorb a finite loop contribution into the property of
the constituent quark
\begin{equation}
\mu_p = {4 \over 3} \, \left ( \mu_u + \mu_l \right )
      - {1 \over 3} \, \left ( \mu_d + \mu_l \right ) \, ,
\end{equation}
and still correctly describe the neutron by an exchange of $u$ and $d$
quarks as
\begin{equation}
\mu_n = {4 \over 3} \, \left ( \mu_d + \mu_l \right )
      - {1 \over 3} \, \left ( \mu_u + \mu_l \right ) \, .
\end{equation}
Of course, the absorption of $\mu_l$ into the constituent quark
property breaks isospin symmetry.  Since most constituent quark models
of octet baryons do break isospin symmetry, the physics of the
$\overline q q$ sea is included in the constituent quark.  Thus, the
simple quark model always did predict some strangeness in the nucleon.
As we will see, the amount may be determined through a measure of
isospin symmetry violation and ratios of light to strange constituent
quark masses.  Since constituent quarks in octet baryons already have
the physics of sea-quarks included intrinsicly, the concept of a
Fock-space expansion of $qqq(\overline q q)^n$ constituent quarks
seems redundant.

   The connected insertions of the current denoted by $L$ or $H$ in
(\ref{decupletEq}) are symmetric within an isospin multiplet.  Hence
taking octet baryon properties with loop contributions absorbed into
the effective degrees of freedom and applying them to decuplet baryon
properties \cite{linde95} violates the QCD equalities among connected
insertions of the probing current.

\subsection{Chiral Perturbation Theory}

   The effective Lagrangian approach of chiral perturbation theory
($\chi$PT) describes baryons as an octet-meson cloud of approximate
Goldstone bosons coupled to nonrelativistic baryons.  Despite the
phenomenological need for the introduction of an explicit $\Delta$
resonance \cite{jenkins91}, as well as the Roper multiplet
\cite{banerjee95}, the approach is argued to capture the essence of
nonperturbative QCD.

   Recently, relationships among the decuplet baryon magnetic moments
were established in a rigorous one-loop, ${\cal O}(p^2)$, $\chi$PT
analysis \cite{banerjee95} where careful regard was given to the
renormalizability of the approach.  The relationships among the
magnetic moments are \cite{banerjee95}
\begin{mathletters}
\begin{eqnarray}
\Delta^{++}  &=& -2\, \Omega^-  - 3\, \Delta^0 \, , \\
\Delta^{+}   &=& -1\, \Omega^-  - 1\, \Delta^0 \, , \\
\Delta^{0}   &=& +0\, \Omega^-  + 1\, \Delta^0 \, , \\
\Delta^{-}   &=& +1\, \Omega^-  + 3\, \Delta^0 \, , \\
\Sigma^{*+}  &=& -1\, \Omega^-  - 2\, \Delta^0 \, , \\
\Sigma^{*0}  &=& +0\, \Omega^-  + 0\, \Delta^0 \, , \\
\Sigma^{*-}  &=& +1\, \Omega^-  + 2\, \Delta^0 \, , \\
\Xi^{*0}     &=& +0\, \Omega^-  - 1\, \Delta^0 \, , \\
\Xi^{*-}     &=& +1\, \Omega^-  + 1\, \Delta^0 \, , \\
\Omega^-     &=& +1\, \Omega^-  + 0\, \Delta^0 \, .
\end{eqnarray}%
\label{chiPTcrypt}%
\end{mathletters}%
At first glance, it would appear that (\ref{chiPTcrypt}) fails to
follow the QCD equalities.  Moreover, it appears on the surface that
even the simple physics of SU(3) flavor symmetry breaking in the
valence sector is absent in (\ref{chiPTcrypt}).  However, demanding
these relationships among decuplet baryon magnetic moments satisfy the
QCD equalities of (\ref{decupletEq}) establishes relationships between
SU(3) breaking in the disconnected loop sector and SU(3) breaking in
the valence sector.  The following relationships are consistent with
$\chi$PT to one-loop order.

   The $\chi$PT results satisfy the QCD equalities by allowing the
$\Delta^0$ moment, reflecting the breaking of SU(3) flavor in the
disconnected sea contribution, to introduce SU(3) flavor symmetry
breaking into the valence sector.  Equating the two representations
provides the following particularly interesting relation
\begin{equation}
L_\Delta = L_\Sigma = L_\Xi = - \Omega^- - 2\, \Delta^0 \, .
\end{equation}
This relation indicates $\chi$PT has failed to resolve any environment
sensitivity in the light quark sector contributions to magnetic
moments.  This is truly disappointing, as this is where the
interesting physics resides in decuplet baryon properties.  The only
consolation is that lattice QCD calculations \cite{leinweber92b}
suggest that such an approximation is reasonable in decuplet baryons,
whereas it is not for octet baryons.  However, the symmetries of
$\chi$PT to this order are not the symmetries of QCD.

   In the $\Sigma$ and $\Xi$ systems, there are two linearly
independent equations and three unknown quark sector contributions.
While we are fortunate to isolate the light valence sector, it is not
possible to simultaneously ascertain the environment independence of
both $H$ and $O$ contributions.  In any event, the $\chi$PT results
are in accord with the following relations
\begin{mathletters}
\begin{eqnarray}
\Delta^{++} &=& 3\, \mu_u + \mu_l \, , \\
\Delta^{+}  &=& 2\, \mu_u + \mu_d + \mu_l \, , \\
\Delta^{0}  &=& \mu_u + 2\, \mu_d + \mu_l \, , \\
\Delta^{-}  &=& 3\, \mu_d + \mu_l \, , \\
\Sigma^{*+} &=& 2\, \mu_u + \mu_s + \mu_l \, , \\
\Sigma^{*0} &=& \mu_u + \mu_d + \mu_s + \mu_l \, , \\
\Sigma^{*-} &=& 2\, \mu_d + \mu_s + \mu_l \, , \\
\Xi^{*0}    &=& 2\, \mu_s + \mu_u + \mu_l \, , \\
\Xi^{*-}    &=& 2\, \mu_s + \mu_d + \mu_l \, , \\
\Omega^{-}  &=& 3\, \mu_s + \mu_l \, ,
\end{eqnarray}%
\end{mathletters}%
where,
\begin{mathletters}
\begin{eqnarray}
\mu_u &=&  {2 \over 3} \, L = {2 \over 3} \left ( - \Omega^- - 2 \,
                             \Delta^0 \right ) \, , \\
\mu_d &=& -{1 \over 3} \, L = -{1 \over 3} \left ( - \Omega^- - 2 \,
                             \Delta^0 \right ) \, , \\
\mu_s &=& -{1 \over 3} \, H = -{1 \over 3} \left ( - \Omega^- + \,
                             \Delta^0 \right ) \, , \\
\mu_l &=& \Delta^0 \, .
\end{eqnarray}%
\end{mathletters}%
Of course this is simply the naive constituent quark model result with
an explicit disconnected sea-quark contribution.  Moreover, the
qualitative agreement between the valence quark sector results of
lattice QCD calculations and the simple quark model has already been
established in \cite{leinweber92b}.

   The new information obtained from $\chi$PT is the explicit
relationship between SU(3) breaking in the disconnected loop and that
in the valence sector.  The vanishing $\Sigma^{*0}$ moment in $\chi$PT
trivially provides the $\Delta^0$ moment as a sum of constituent
quark moments
\begin{mathletters}
\begin{eqnarray}
\Delta^0 &\simeq& - \left ( \mu_u + \mu_d + \mu_s \right ) \, , \\
         &\simeq& - {1 \over 3} \left ( L - H \right ) \, .
\end{eqnarray}%
\label{ValLoopRel}%
\end{mathletters}%
The valence sectors of $\Sigma^{*0}$ and $\Xi^{*0}$ may be used to
estimate the $\Delta^0$ magnetic moment.  The lattice QCD calculations
of Ref.\ \cite{leinweber92b} indicate
\begin{mathletters}
\begin{eqnarray}
\Delta^0 &=& - \Sigma^{*0}_{\rm Val} = -0.29 \pm 0.05\ \mu_N \, , \\
         &=& - {1 \over 2} \, \Xi^{*0}_{\rm Val} = -0.33 \pm 0.07\
         \mu_N \, ,
\end{eqnarray}%
\end{mathletters}%
and the weighted average of these results is
\begin{equation}
\mu_l^\Delta = \Delta^0 = -0.30 \pm 0.04\ \mu_N \, .
\label{DecLoopEst}
\end{equation}
This result is approximately three times the result obtained for octet
baryons under the assumption of no environment sensitivity for the
disconnected sea-quark loop \cite{leinweber92a}.  There it was found
\begin{equation}
\mu_l^N = -0.10 \pm 0.06\ \mu_N \, .
\label{OctLoopEst}
\end{equation}
Naively, one might expect such a factor of three as the spin of the
$\Delta$ is three times the spin of the nucleon.

   The disconnected sea-quark loop contribution of (\ref{DecLoopEst})
applied to the $\Omega^-$ is precisely the contribution required to
augment the earlier reported \cite{leinweber92b} valence sector
contribution of $-1.73(22)\ \mu_N$ to $-2.03(22)\ \mu_N$ and restore
agreement with the new precision measurement of the $\Omega^-$
magnetic moment \cite{wallace95} of $-2.024(56)\ \mu_N$.

   The disconnected sea-quark loop contribution to decuplet baryon
magnetic moments is not small.  Given the recent flurry of activity
surrounding decuplet baryon magnetic moments, a summary of the
predictions for these moments obtained from lattice QCD
\cite{leinweber91,leinweber92b} including the disconnected sea
contributions is provided in Table \ref{MagMomPred}.  The chief
assumption in obtaining the loop contribution for decuplet baryons is
the independence of loop contributions from environment effects, an
assumption supported by the agreement of estimates for $\mu_l^\Delta$
{}from $\Sigma^{*0}$ and $\Xi^{*0}$.

\begin{table}[t]
\caption{Predictions for octet and decuplet baryon magnetic moments
{}from lattice QCD are compared with experimental results
\protect\cite{wallace95,rpp94} where available.  A single scale
parameter accounting for lattice artifacts has been introduced to
reproduce the quantity $\mu_{\Sigma^+} - \mu_{\Sigma^-}$ which is
independent of disconnected sea-quark loops.  Uncertainties in the
last digit(s) are indicated in parentheses.}
\label{MagMomPred}
\setdec 00.0000(00)
\begin{tabular}{lcc}
Baryon        &Lattice QCD       &Experiment \\
              &$(\mu_N)$         &$(\mu_N)$  \\
\tableline
Octet\tablenotemark[1] \\
$p$           &\dec    2.72(26)  &\dec    2.793      \\
$n$           &\dec $-$1.82(34)  &\dec $-$1.913      \\
$\Sigma^{+}$  &\dec    2.47(9)   &\dec    2.458(10)  \\
$\Sigma^{0}$  &\dec    0.66(8)   & \\
$\Sigma^{-}$  &\dec $-$1.15(9)   &\dec $-$1.160(25)  \\
$\Xi^{0}   $  &\dec $-$1.27(14)  &\dec $-$1.250(14)  \\
$\Xi^{-}   $  &\dec $-$0.63(7)   &\dec $-$0.6507(25) \\
\\
Decuplet\\
$\Delta^{++}$ &\dec    5.98(93)  &\dec 3.7$\to 7.5$ \\
$\Delta^{+} $ &\dec    2.83(45)  \\
$\Delta^{0} $ &\dec $-$0.319(53) \\
$\Delta^{-} $ &\dec $-$3.47(52)  \\
$\Sigma^{*+}$ &\dec    2.95(33)  \\
$\Sigma^{*0}$ &\dec    0.021(12) \\
$\Sigma^{*-}$ &\dec $-$2.91(33)  \\
$\Xi^{*0}   $ &\dec    0.276(39) \\
$\Xi^{*-}   $ &\dec $-$2.47(24)  \\
$\Omega^{-} $ &\dec $-$2.11(20)  &\dec $-$2.024(56) \\
\end{tabular}
\tablenotetext[1]{$\Lambda^0$ is omitted as the disconnected sea-quark
loop contribution cannot be estimated from the QCD equalities. }
\end{table}

   The scale parameter accounting for lattice artifacts is selected to
reproduce the quantity $\mu_{\Sigma^+} - \mu_{\Sigma^-}$ which is
independent of disconnected sea-quark loops and provides minimal
statistical uncertainties in the predictions.  This scale parameter
differs slightly from the parameter reproducing the proton moment
used up to this point.  The $\Delta^0$ moment indicates the best
estimate for $\mu_l^\Delta = -0.319 \pm 0.053$.

   Octet baryon moments with the disconnected sea-quark loop
contributions of (\ref{OctetLoops}) are also indicated in Table
\ref{MagMomPred} to provide some reference on the reliability of the
predictions.  It is remarkable that a single scale parameter applied
to the lattice QCD results is all that is required to reproduce the
octet baryon magnetic moments within uncertainties.  The essential
ingredients to any model which hopes to reproduce these moments are an
environment sensitivity of the quark sector contributions and a small
but essential disconnected sea-quark contribution.

\section{Strangeness Contributions to Nucleon Moments}

   The idea that the contributions of various flavors running around
the disconnected sea loop might be estimated by considering ratios of
constituent quark masses is not new \cite{karl92}.  However, the
connection between quarks running around the disconnected loop and
constituent quarks is obscured by the fact that constituent quarks are
usually associated with the valence quark sector carrying the quantum
numbers of the hadron.  However, $\chi$PT and the QCD equalities
presented here have provided a quantitative link between the two,
consistent to ${\cal O}(p^2)$ and one-loop in $\chi$PT.

   With this new relationship, the strangeness contribution to nucleon
moments may be isolated.  Denoting $\mu_l^u$ as the $u$-quark
contribution to the disconnected sea loop in the nucleon, etc.
\begin{mathletters}
\begin{eqnarray}
\mu_l^N &=& \mu_l^u + \mu_l^d + \mu_l^s \, , \\
        &=& - \left ( {\mu_{\rm Val}^d \over \mu_{\rm Val}^s} - 1
              \right ) \, \mu_l^s \, .
\end{eqnarray}%
\end{mathletters}%
Here, the equivalence of light to heavy magnetic moment ratios in the
valence and loop sector has been assumed, in accord with the
discussion surrounding (\ref{ValLoopRel}).  The doubly represented
quark sector contributions have smaller statistical uncertainties and
are more like constituent quarks than the singly represented quark
sector \cite{leinweber92b}.  Using the doubly represented quark
sectors of $\Sigma$ and $\Xi$ discussed in Section \ref{OctetSym} to
estimate the ratio of quark moments
\begin{eqnarray}
\mu_l^s &=& - \, { D_\Xi / D_\Sigma \over
                   \left ( 1 -  D_\Xi / D_\Sigma \right ) } \,
              \mu_l^N \, ,
\label{NuclStrangeMom} \\
        &=& +0.25 \pm 0.10\ \mu_N \, . \nonumber
\end{eqnarray}
Table \ref{GMs} compares this result with other estimates for the
strangeness contribution to nucleon magnetic moments.  This new result
is large relative to other estimates.

\begin{table}[t]
\caption{Various estimates for strange quark contributions to nucleon
         magnetic moments.}
\label{GMs}
\setdec 0.000000000
\begin{tabular}{llc}
Approach    &Reference                           &$G_M^s(0) = -3
\mu_l^s$ \\
            &                                    &$(\mu_N)$
            \\
\tableline
QCD Equalities &This work                        &\dec $-$0.75$\pm
0.30$ \\
Lattice QCD &Leinweber \cite{leinweber95i,leinweber92a}
                                                 &\dec $-$0.33 \\
Poles       &Hammer {\it et al.} \cite{hammer95} &\dec $-$0.24$\pm
0.03$ \\
SU(3) NJL   &Kim {\it et al.} \cite{kim95}       &\dec $-$0.45 \\
Kaon Loops  &Musolf and Burkardt \cite{musolf94} &\dec $-$0.31$\to
-0.40$ \\
Kaon Loops  &Cohen {\it et al.} \cite{cohen93}   &\dec $-$0.24$\to
-0.32$ \\
Vector Dom. &Cohen {\it et al.} \cite{cohen93}   &\dec $-$0.24$\to
-0.32$ \\
SU(3) Skyrme&Park {\it et al.} \cite{park91}     &\dec $-$0.13 \\
Poles       &Jaffe \cite{jaffe89}                &\dec $-$0.31$\pm
0.09$ \\
Simple Quark Model &This work and Ref.\ \cite{rpp94}
                                                 &\dec $-$0.20 \\
\end{tabular}
\end{table}

The large value found here relative to that in Ref.
\cite{leinweber95i} reflects the environment sensitivity of the
disconnected sea-quark loop contribution and the ratio of strange to
light quark contributions assumed in resolving the strangeness
contribution.  In Ref. \cite{leinweber95i} $D_\Xi / D_\Sigma$ was
crudely estimated to be 1/2.  Using the value of (\ref{MagMomRat})
provides $G_M^s(0) = -0.42(25)\ \mu_N$ assuming environment
independence of the disconnected sea-quark loop contributions.  This
correction is small relative to that obtained by allowing for
environment effects in the extraction of loop contributions.

In section IV we established that constituent quark models which break
isospin symmetry between the $u$ and $d$ quark moments have the
physics of the disconnected sea-quark loops included intrinsicly.
Taking the usual fit of $\mu_u^F$, $\mu_d^F$, and $\mu_s^F$ intrinsic
quark moments to $p$, $n$, and $\Lambda$ baryon moments \cite{rpp94},
the disconnected sea-quark loop contribution may be isolated from
\begin{eqnarray}
\mu_l^N &=& {1 \over 3} \left ( 2\, \mu_d^F + \mu_u^F \right ) \, \\
        &=& -0.031\ \mu_N \, . \nonumber
\end{eqnarray}
This result is small enough that one might begin to worry about
isospin violation in the current quark masses.  Accounting for a five
MeV current quark mass difference, $m_d - m_u$, increases this result
to $\sim -0.04\ \mu_N$.  Application of (\ref{NuclStrangeMom}) using
$D_\Xi / D_\Sigma = \mu_s^F/\mu_d^F$, provides the simple quark model
prediction of the strangeness contribution to nucleon magnetic moments
indicated in Table \ref{GMs}.

\begin{table}[t]
\caption{Predictions for the quark sector contributions to baryon
magnetic moments.  Quark charges are included, such that the row sum
reproduces the baryon moment.  A single scale parameter accounting for
lattice artifacts has been introduced to reproduce the quantity
$\mu_{\Sigma^+} - \mu_{\Sigma^-}$.  Uncertainties in the last digit(s)
are indicated in parentheses.  All moments are in units of $\mu_N$.}
\label{QrkMomPred}
\setdec 00.000(00)
\begin{tabular}{lcccccc}
Baryon        &\multicolumn{3}{c}{Valence Sector}
                              &\multicolumn{3}{c}{Disconnected Loop
                                Sector} \\
              &$u$    &$d$    &$s$   &$u$    &$d$    &$s$       \\
\tableline
$p$           &\dec 2.74(20)  &\dec 0.15(11)  &\dec 0.
              &\dec $-$0.84(24)  &\dec 0.42(12)  &\dec 0.25(10)   \\
$n$           &\dec $-$0.30(21)  &\dec $-$1.37(10)  &\dec 0.
              &\dec $-$0.84(24)  &\dec 0.42(12)  &\dec 0.25(10)   \\
$\Sigma^+$    &\dec 2.42(4)   &\dec 0.        &\dec 0.08(5)
              &\dec $-$0.15(20)  &\dec 0.07(10)  &\dec 0.04(7)   \\
$\Sigma^0$    &\dec 1.21(2)   &\dec $-$0.60(1)   &\dec 0.08(5)
              &\dec $-$0.15(20)  &\dec 0.07(10)  &\dec 0.04(7)   \\
$\Sigma^-$    &\dec 0.        &\dec $-$1.21(2)   &\dec 0.08(5)
              &\dec $-$0.15(20)  &\dec 0.07(10)  &\dec 0.04(7)   \\
$\Xi^0$       &\dec $-$0.43(9)   &\dec 0.        &\dec $-$0.71(5)
              &\dec $-$0.67(18)  &\dec 0.34(9)   &\dec 0.20(7)   \\
$\Xi^-$       &\dec 0.           &\dec 0.22(5)   &\dec $-$0.71(5)
              &\dec $-$0.67(18)  &\dec 0.34(9)   &\dec 0.20(7)   \\
$\Delta^{++}$ &\dec 6.30(96)     &\dec 0.           &\dec 0.
              &\dec $-$1.54(21)  &\dec 0.77(10)     &\dec 0.45(9)   \\
$\Delta^+$    &\dec 4.20(64)     &\dec $-$1.05(16)  &\dec 0.
              &\dec $-$1.54(21)  &\dec 0.77(10)     &\dec 0.45(9)   \\
$\Delta^0$    &\dec 2.10(32)     &\dec $-$2.10(32)  &\dec 0.
              &\dec $-$1.54(21)  &\dec 0.77(10)     &\dec 0.45(9)   \\
$\Delta^-$    &\dec 0.           &\dec $-$3.15(48)  &\dec 0.
              &\dec $-$1.54(21)  &\dec 0.77(10)     &\dec 0.45(9)   \\
$\Sigma^{*+}$ &\dec 3.90(44)     &\dec 0.           &\dec $-$0.64(7)
              &\dec $-$1.54(21)  &\dec 0.77(10)     &\dec 0.45(9)   \\
$\Sigma^{*0}$ &\dec 1.95(22)     &\dec $-$0.98(11)  &\dec $-$0.64(7)
              &\dec $-$1.54(21)  &\dec 0.77(10)     &\dec 0.45(9)   \\
$\Sigma^{*-}$ &\dec 0.           &\dec $-$1.96(22)  &\dec $-$0.64(7)
              &\dec $-$1.54(21)  &\dec 0.77(10)     &\dec 0.45(9)   \\
$\Xi^{*0}$    &\dec 1.83(18)     &\dec 0.           &\dec $-$1.24(5)
              &\dec $-$1.54(21)  &\dec 0.77(10)     &\dec 0.45(9)   \\
$\Xi^{*-}$    &\dec 0.           &\dec $-$0.91(9)   &\dec $-$1.24(5)
              &\dec $-$1.54(21)  &\dec 0.77(10)     &\dec 0.45(9)   \\
$\Omega^-$    &\dec 0.           &\dec 0.           &\dec $-$1.79(16)
              &\dec $-$1.54(21)  &\dec 0.77(10)     &\dec 0.45(9)   \\
\end{tabular}
\end{table}

Table \ref{QrkMomPred} provides a breakdown of the various quark
sector contributions to baryon magnetic moments.  It will be
interesting to confront the predictions for the nucleon with the
experimental determinations anticipated from CEBAF.  The sector
contributions $u:d:s$ of $1.90(30) : 0.57(16) : 0.25(10)$ for the
proton are sufficiently different from the traditionally viewed simple
quark model contributions of $2.47 : 0.32 : 0$ to be interesting.
Similarly, the neutron sector contributions are $-1.14(32) : -0.95(16)
: 0.25(10)$ to be compared with $-1.30 : -0.62 : 0$ in the traditional
simple quark model.  It is interesting to note that when isospin
symmetry breaking in the $u$-$d$ moments is used to isolate the
strangeness contribution in the simple quark model, the quark sector
contributions of $u:d:s = 2.30 : 0.42 : 0.07$ for the proton and
$-1.13 : -0.85 : 0.07$ for the neutron look more similar to the
predictions from lattice QCD and QCD equalities.

\section{Summary}

   Using valence sector information coupled with experiment where
available, or chiral perturbation theory, we have seen how QCD
equalities can be used to gain insight into sea-quark contributions,
and in particular, strangeness contributions to nucleon properties.
The QCD equalities were derived for a general quark current, bilinear
in the quark fields.

Here, the focus in practice has been on the magnetic properties of
baryons.  In particular, the disconnected sea-quark loop contribution
to the nucleon moment is determined to be $-0.17 \pm 0.07\ \mu_N$, and
the strange quark contributes $+0.25 \pm 0.10\ \mu_N$ to the loop.
For decuplet baryons, the disconnected sea-quark loop contribution is
estimated to be $-0.32 \pm 0.05\ \mu_N$, of which $+0.45 \pm 0.09\
\mu_N$ has its origin from the strange quark.

The QCD equalities have also resolved the equivalence of $\chi$PT to
${\cal O}(p^2)$ and the simple quark model with an explicit
disconnected sea-quark contribution.  New relationships between
SU(3)-flavor symmetry breaking in valence and sea contributions were
obtained.  Moreover, an implicit strangeness contribution was
identified in simple constituent quark models where isospin symmetry
between the $u$ and $d$-quark moments, $\mu_u = -2 \, \mu_d$, is
broken.

The techniques demonstrated here may be applied to other observables
of quark current operators.  For example, the strangeness radius of
the nucleon might be extracted from a study of experimentally measured
nucleon form factors and models of $\Xi$ and $\Sigma$ form factors.
Favorable combinations could be found which minimize the model
dependence.

   Finally, it is hoped that these equalities will be useful in the
development of more sophisticated models of QCD.  As one considers the
possible refinements of quark models, it is important to maintain the
symmetries of QCD.  Not all QCD-inspired models have succeeded in
doing this.

\acknowledgements

Thanks to Joe Milana for his interest in making comparisons between
the expectations of chiral perturbation theory and lattice QCD
calculations, and for stimulating my interest in defining rigorous
relationships among these observables.  This research was supported by
the U.S. Department of Energy under grant DE-FG06-88ER40427.

\end{document}